\documentclass[10pt,conference]{IEEEtran}
\IEEEoverridecommandlockouts

\usepackage{tcolorbox}
\usepackage{fbox}
\usepackage{xcolor}
\usepackage{booktabs}
\usepackage{verbatim}
\usepackage{pgfplots}
\usepackage{hyperref}
\usepackage{filecontents}
\usepackage{tikz}
\usepackage{subfigure}
\usepackage{graphics}
\usepackage{multirow}
\usepackage{fancybox}
\usepackage[switch]{lineno}
\usepackage{balance}
\usepackage[flushleft]{threeparttable}
\usetikzlibrary{patterns}

\pgfplotsset{compat=1.16}

\definecolor{red}{rgb}{1, 0, 0}
\definecolor{green}{rgb}{0, 1, 0}
\definecolor{blue}{rgb}{0, 0, 1}

\newcommand\tool{NeuralLog }
\newcommand{\repourl}{https://github.com/vanhoanglepsa/NeuralLog}

\begin{document}
\title{Log-based Anomaly Detection Without Log Parsing
}



\author{\IEEEauthorblockN{
Van-Hoang Le and
Hongyu Zhang\IEEEauthorrefmark{2}\thanks{\IEEEauthorrefmark{2}Hongyu Zhang is the corresponding author.}
\IEEEauthorblockA{The University of Newcastle, NSW, Australia}
\IEEEauthorblockA{vanhoang.le@uon.edu.au, hongyu.zhang@newcastle.edu.au}
}\\\vspace{-30pt}}





\maketitle

\begin{abstract}
Software systems often record important runtime information in system logs for troubleshooting purposes. 
There have been many studies that use log data to construct machine learning models for detecting system anomalies. Through our empirical study, we find that existing log-based anomaly detection approaches are significantly affected by log parsing errors that are introduced by 1) OOV (out-of-vocabulary) words, and 2) semantic misunderstandings. The log parsing errors could cause the loss of important information for anomaly detection. To address the limitations of existing methods, we propose NeuralLog, a novel log-based anomaly detection approach that does not require log parsing. NeuralLog extracts the semantic meaning of raw log messages and represents them as semantic vectors. These representation vectors are then used to detect anomalies through a Transformer-based classification model, which can capture the contextual information from log sequences. Our experimental results show that the proposed approach can effectively understand the semantic meaning of log messages and achieve accurate anomaly detection results. Overall, NeuralLog achieves F1-scores greater than 0.95 on four public datasets, outperforming the existing approaches.
\end{abstract}


\vspace{-1mm}
\section{Introduction}
\label{sec:introduction}
High availability and reliability are essential for large-scale software-intensive systems \cite{bauer2012reliability, kazemzadeh2009reliable}. With the increasing complexity and scale of systems, anomalies have become inevitable. A small problem in the system could lead to performance degradation, data corruption, and even a significant loss of customers and revenue. Anomaly detection is, therefore, necessary for the quality assurance of complex software-intensive systems.

Software-intensive systems often generate console logs to record system states and critical events at runtime. Engineers can utilize log data to understand the system status, detect the anomalies, and identify the root causes. As the amount of logs could be huge, anomaly detection based on manual analysis of logs is time-consuming and error-prone. Over the years, many data-driven methods have been proposed to automatically detect anomalies by analyzing log data \cite{breier2015anomaly, zhang2020anomaly, du2017deeplog, chen2004failure, xu2009detecting, zhang2019robust, guo2021logbert}. Machine learning-based methods (such as Logistic Regression \cite{bodik2010fingerprinting}, Support Vector Machine \cite{chen2004failure}, Invariant Mining \cite{lou2010mining}) extract log events and adopt supervised or unsupervised learning to detect the occurrences of system anomalies.

Recently, some deep learning-based approaches have been proposed. For example, LogRobust \cite{zhang2019robust} and LogAnomaly \cite{meng2019loganomaly} adopt Word2vec model \cite{mikolov2013efficient} to obtain embedding vectors of log events, then applied an LSTM model to detect anomalies.

However, the existing approaches rely on log parsing to preprocess semi-structured log data. Log parsers remove the variable part from log messages and retain the constant part to obtain log events.
To investigate the inaccuracy of log parsing, we have performed an empirical study on real-world log data. We find that existing log parsers produce a noticeable number of parsing errors, which directly downgrade anomaly detection performance. 
The log parsing errors are mainly due to the following two reasons: 
1) The logging statements could frequently change during software evolution, resulting in new log events that were not seen in training data;
2) Valuable information could be lost while parsing log messages into log events, which may lead to misunderstanding of the semantic meaning of log messages.
Our empirical study also finds that the log parsing errors can affect the follow-up anomaly detection task and decrease the detection accuracy.

To overcome the above-mentioned limitations of existing approaches, we propose NeuralLog, a novel anomaly detection approach, which can achieve effective and efficient anomaly detection on real-world datasets. Unlike the existing approaches, \tool does not rely on any log parsing, thus preventing the loss of information due to log parsing errors. Each log message is directly transformed into a semantic vector, which is capable of capturing both semantic information embedded in log messages and the relationship between log messages. Then, taking a sequence of semantic vectors as input, a Transformer-based classification model is applied to detect anomalies. The Transformer-based model with multi-head self-attention mechanism \cite{vaswani2017attention} can learn the contextual information from the log sequences in the form of vector representations. As a result, \tool is effective for log-based anomaly detection.

We have evaluated the proposed approach using four public datasets. The experimental results show that \tool can understand the semantic meaning of log data and adequately handle OOV words. 
It achieves high F1-scores (all greater than 0.95) for anomaly detection and outperforms the existing log-based anomaly detection approaches. 

The main contributions of this paper are as follows:
\begin{enumerate}
\item We perform an empirical study of log parsing errors. We find that existing log-based anomaly detection approaches are adversely affected by the log parsing errors introduced by the OOV words and semantic misunderstanding.
\item We propose NeuralLog, a novel deep learning-based approach that can detect system anomalies without log parsing. \tool utilizes BERT, a widely-used pre-trained language representation, to encode the semantic meaning of log messages.
\item We have evaluated \tool using public datasets. The results confirm the effectiveness of \tool for representing log messages and detecting anomalies.
\end{enumerate}


\section{Background}
\label{sec:motivation}
\subsection{Log Data}
\label{sec:background_log_data}
Large and complex software-intensive systems often produce a large amount of log data for troubleshooting purposes during system operation. Log data records the system's events and internal states during runtime. By analyzing logs, operators can better understand systems' status and diagnose the system when a failure occurs.

\begin{figure}[h]
    \vspace{-1mm}
    \centering
    \includegraphics[width=.85\linewidth]{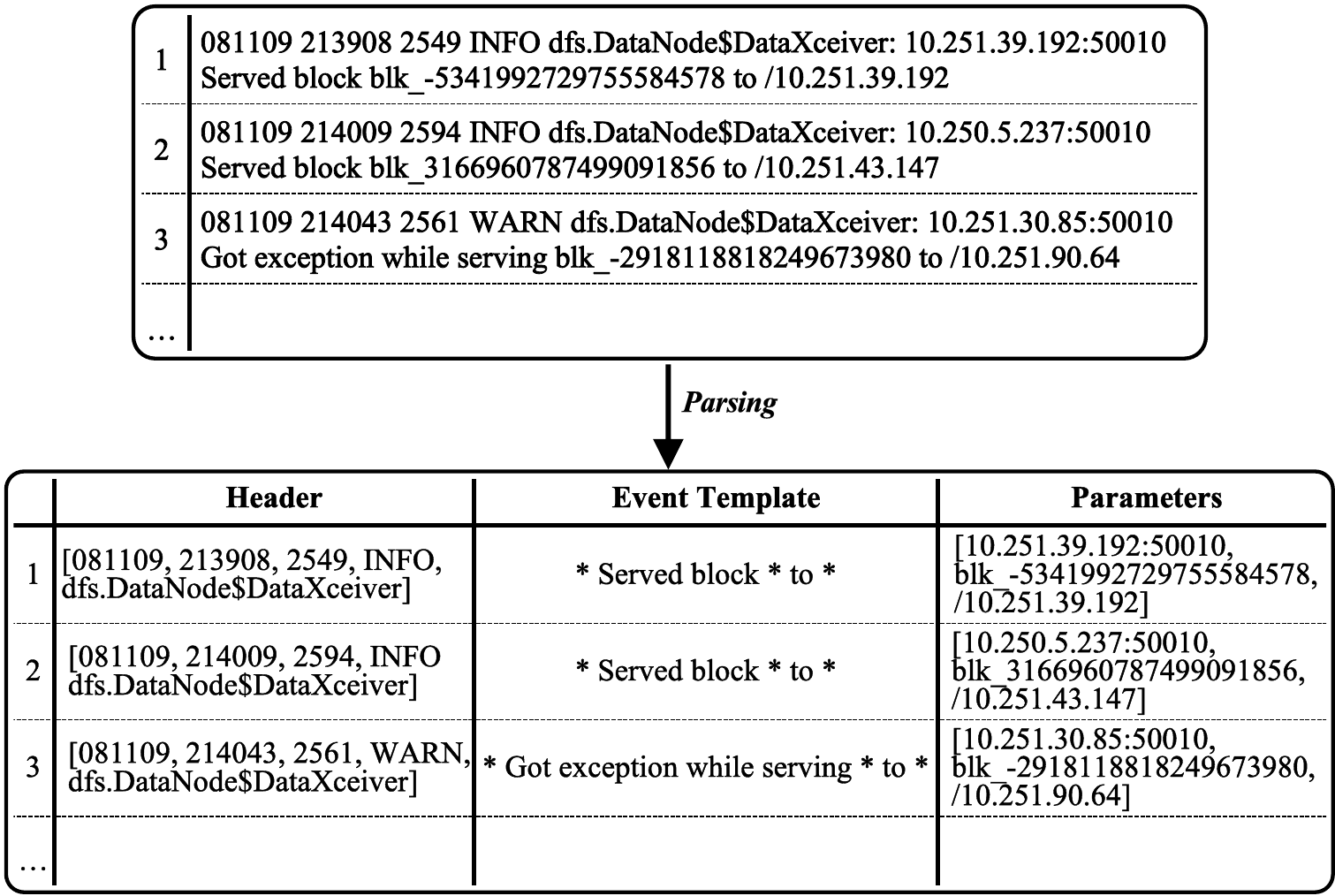}
    \caption{An Example of HDFS Logs and Parsed Results}
    \label{fig:hdfs_log}
    \vspace{-1mm}
\end{figure}

Figure \ref{fig:hdfs_log} shows a snippet of raw logs generated by HDFS (Hadoop Distributed File System). The raw log messages are semi-structured texts, which contain \textit{header} and \textit{content}.
The header is determined by the logging framework 
and includes information such as timestamp, verbosity level (e.g., WARN/INFO), and component \cite{zhu2019tools}. The log content consists of a constant part (keywords that reveal the event template) and a variable part (parameters that carry dynamic runtime information).
Log parsing automatically converts each log message into a specific event template by removing parameters and keeping the keywords.
The log events can be grouped into \textit{log sequences} (i.e., series of log events that record specific execution flows) according to sessions or fixed/sliding time windows \cite{he2016evaluation}. 


\subsection{Log Parsing Methods}
\label{sec:background_log_parsing}
Log parsing automatically converts each log message into a specific event template by removing parameters and keeping the keywords.
For example, the log template \textit{``$*$ Served block $*$ to $*$"} can be extracted from the first log message in Figure \ref{fig:hdfs_log}. Here, \textit{``$*$"} denotes the position of a parameter.

There are many log parsing techniques, including frequent pattern mining \cite{nagappan2010abstracting, vaarandi2015logcluster, dai2020logram}, clustering \cite{tang2011logsig, hamooni2016logmine, shima2016length}, language modeling \cite{thaler2017towards}, heuristics \cite{he2017drain, jiang2008abstracting, makanju2009clustering}, etc.
The heuristics-based approaches make use of the characteristics of logs and have been found to perform better than other techniques in terms of accuracy and time efficiency \cite{zhu2019tools}.
In this study, we evaluate four top-ranked parsers include Drain \cite{he2017drain}, AEL \cite{jiang2008abstracting}, IPLoM \cite{makanju2009clustering} and Spell \cite{du2016spell}.
They utilize the characteristics of tokens (e.g., occurrences, position, relation, etc.) and special structures (e.g., a tree) to represent log messages and extract common templates.
Drain applies a fixed-depth tree structure to represent log messages and extracts common templates effectively. Spell utilizes the longest common subsequence algorithm to parse logs in a stream manner.
AEL separates log messages into multiple groups by comparing the occurrences between constants and variables. IPLoM employs an iterative partitioning strategy, which partitions log messages into groups by message length, token position, and mapping relation.
These log parsers are widely used in existing studies and have proven their efficiency on real-world datasets \cite{zhu2019tools, he2016evaluation,he2016experience, el2020systematic}.

\subsection{Log-based Anomaly Detection}
\label{sec:log-based-anomaly-detection}
\begin{figure}[h]
    \vspace{-1mm}
    \centering
    \includegraphics[width=0.8\linewidth]{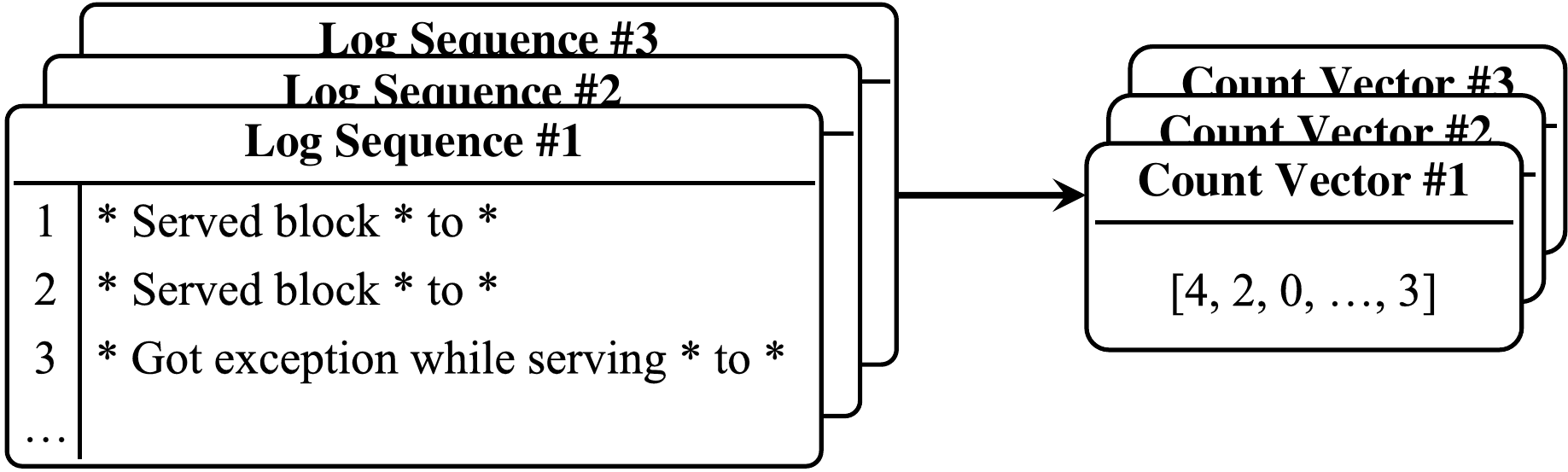}
    \caption{An Example of Log Sequence, and Log Count Vector}
    \label{fig:hdfs_log_sequence_and_vector}
    \vspace{-1mm}
\end{figure}

Over the years, many log-based anomaly detection approaches have been proposed. Some of them are based on unsupervised learning methods, which require only unlabeled data to train. For example, Xu et al. \cite{xu2009detecting} employed Principal Component Analysis (PCA) to generate two subspaces (normal space and anomaly space) of log count vectors.
If a log sequence has its log count vector far from the normal space, it is considered an anomaly. IM \cite{lou2010mining} and ADR \cite{zhang2020anomaly} discover the linear relationships among log events from log count vectors. Those log sequences that violate the relationship are considered anomalies. 
There are also many supervised anomaly detection approaches. For example, \cite{liang2007failure, chen2004failure, bodik2010fingerprinting} represent log sequences as log count vectors, then applied Support Vector Machine (SVM), Logistic Regression (LR), and Decision Tree algorithm to detect anomalies, respectively.
These approaches have many common characteristics. They all require a log parser to preprocess and extract log templates from log messages. Then, the occurrences of log templates are
counted, resulting in log count vectors. Finally, a machine learning model is constructed to detect anomalies. 
Figure \ref{fig:hdfs_log_sequence_and_vector} shows an example of log sequence and log count vectors from log templatesproduced within Drain \cite{he2017drain}.


In recent years, many deep learning-based models have been proposed to analyze log data and detect anomalies \cite{du2017deeplog, meng2019loganomaly, zhang2019robust}. For example, DeepLog \cite{du2017deeplog} first applies the Spell \cite{du2016spell} parser to extract log templates. Then, it leverages the indexes of log templates and inputs them to an LSTM model to predict the next log templates. Finally, DeepLog detects anomalies by determining whether or not the incoming log templates are unexpected.
LogAnomaly \cite{meng2019loganomaly} uses log count vector to detect the anomalies reflected by anomalous log event numbers. It proposes a synonyms and antonyms based method to represent the words in log templates.
LogRobust \cite{zhang2019robust} incorporates a pre-trained Word2vec model, namely FastText \cite{joulin2016fasttext}, and combines with TF-IDF weight \cite{salton1988term} for learning the representation vectors of log templates, which are generated by Drain \cite{he2017drain}. Then, these vectors input an Attention-based Bi-LSTM model to detect anomalies.
Due to the imperfection of log parsing, the above methods tend to the lose semantic meaning of log messages, thus leading to inaccurate detection results.

\section{An Empirical Study of Log Parsing Errors}
\label{sec:empirical}
In this section, we describe an empirical study on the problem of existing log parsers and their impact on log-based anomaly detection.
We use two public datasets in our study, namely Blue Gene/L (BGL) \cite{oliner2007supercomputers, he2020loghub} and Thunderbird \cite{oliner2007supercomputers, he2020loghub}.
The datasets are collected between 2004 and 2006 from real-world supercomputing systems \cite{oliner2007supercomputers}
and consist of 14,672,653 log messages in total, among which 353,394 log messages are manually labeled as anomalies. 
\subsection{Log Parsing Errors Introduced by OOV Words}
\label{sec:oov_words}


During development and maintenance, developers can add new log statements to source code and modify the content of existing log statements.
Besides, runtime information can be added to log messages as parameters to record system status.
As a consequence, new words, i.e., out-of-vocabulary (OOV) words, frequently appear in log data. For example, for a log message \textit{``memory manager address \textbf{parity} error"} in BGL, if the word ``parity" did not appear in historical logs, it is an OOV word.
To determine OOV words, we first sort log messages by the timestamps of logs and leverage the front $P$\% (according to the timestamps of logs)
as the training data and the rest as the testing data. Then, we split each training log message into a set of tokens by the whitespace character and build a vocabulary from these tokens.
\textit{OOV words} are those words in testing data that do not exist in the vocabulary.
In this section, we increase the percentages of training data from 20\% to 80\%. Then we calculate the proportion of OOV words in all the splits.

To facilitate understanding, we use the BGL data at the 60/40 splitting (the first 60\% of the BGL dataset is used for training, and the rest is for testing) to explain in detail the analysis of OOV words. The training set contains 153,786 unique words, and the testing set contains 384,730 unique words. Among the unique words in the testing set, 362,123 words (94.12\%) are unseen in the training set. These OOV words only concentrate in a small subset of logs (i.e., 8.51\% of the testing set, which is 160,403 out of 1,885,398 log messages).
These OOV words result in 1,304 unseen log events (i.e., log templates with OOV words) in the testing set, accounting for 86.59\% of the total number of log events in the testing set.

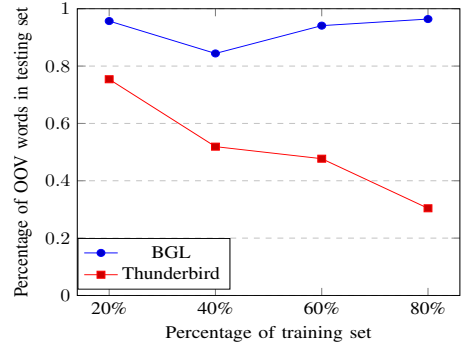
\begin{figure}[h]
\vspace{-1mm}
\centering
\resizebox{0.7\columnwidth}{0.67\height}{
    \begin{tikzpicture}
    \begin{axis}[
        ymax = 1,
        ymin = 0,
        symbolic x coords={20\%, 40\%,60\%,80\%},
        xtick=data,
        ymajorgrids=true,
        grid style=dashed,
        xlabel={Percentage of training set},
        ylabel={Percentage of OOV words in testing set},
        legend style={
            anchor=north east,
            at={(0.4,0.2)}
        }
    ]
    \addplot coordinates {
        (20\%,0.957)  (40\%, 0.844)  (60\%, 0.941)  (80\%, 0.964) 
    };
    \addplot coordinates {
        (20\%,0.754)  (40\%, 0.519)  (60\%, 0.477)  (80\%, 0.304)
    };
    \legend{BGL,Thunderbird}
    \end{axis}
    \end{tikzpicture}
}
\caption{Analysis of OOV words in log messages in public datasets}
\label{fig:oov_percentage}
\vspace{-1mm}
\end{figure}

Figure \ref{fig:oov_percentage} shows the percentages of OOV words in testing data when the percentages of training data increase from 20\% to 80\% on BGL and Thunderbird datasets.
On the BGL dataset, there are always more than 80\% of words in the testing set that are unseen in the training set.
On the Thunderbird dataset, with the growth of training data, the proportion of OOV words in the testing set is gradually decreasing. However, when we use 80\% Thunderbird logs to train the model, we still find 30.4\% of OOV words in the testing set.

\begin{figure}[h]
    \vspace{-1mm}
    \centering
    \subfigure[Log messages with OOV words]{
        \resizebox{0.46\columnwidth}{!}{
            \begin{tikzpicture}
            \centering
            \begin{axis}[
                ymax = 0.3,
                ymin = 0,
                symbolic x coords={20\%, 40\%,60\%,80\%},
                xtick=data,
                ymajorgrids=true,
                grid style=dashed,
                xlabel={Percentage of training set},
                ylabel style = {
                    align=center,
                    text width=6cm
                },
                ylabel=Percentage of log messages with OOV words in testing set
            ]
            \addplot coordinates {
                (20\%, 0.071) (40\%, 0.065)  (60\%, 0.085)  (80\%, 0.067) 
            };
            \addplot coordinates {
                (20\%,0.018)  (40\%, 0.015) (60\%, 0.019)  (80\%, 0.017)
            };
            \legend{BGL,Thunderbird}
            \end{axis}
            \end{tikzpicture}
        }
        \label{fig:message_with_oov}
    }
    \subfigure[Log templates with OOV words] {
        \resizebox{0.46\columnwidth}{!}{
            \begin{tikzpicture}
            \centering
            \begin{axis}[
                ymax = 1,
                ymin = 0,
                symbolic x coords={20\%, 40\%,60\%,80\%},
                xtick=data,
                xlabel={Percentage of training set},
                ylabel style = {
                    align=center,
                    text width=6cm
                },
                ymajorgrids=true,
                grid style=dashed,
                ylabel=Percentage of log templates with OOV words in testing set,
                legend style={
                    anchor=north east,
                    at={(0.4,0.2)}
                }
            ]
            \addplot coordinates {
                (20\%, 0.898) (40\%, 0.873)  (60\%, 0.866)  (80\%, 0.820) 
            };
            \addplot coordinates {
                (20\%,0.878)  (40\%, 0.758) (60\%, 0.676)  (80\%, 0.631)
            };
            \legend{BGL,Thunderbird}
            \end{axis}
            \end{tikzpicture}
        }
        \label{fig:templates_with_oov}
    }
\caption{Analysis of log with OOV words}
\label{fig:oov_percentage_logs}
\vspace{-1mm}
\end{figure}
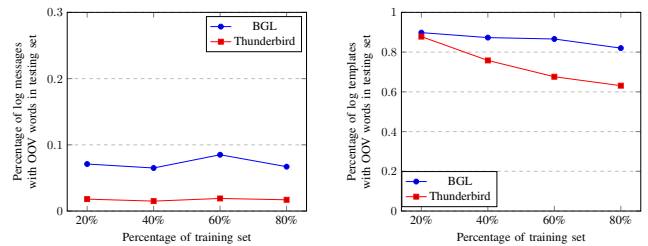

Next, we evaluate the number of log messages (log lines) and log templates that contain OOV words.
The percentage of log messages containing OOV words is shown in Figure \ref{fig:message_with_oov}. It can be seen that the percentage of log messages containing OOV words in the testing set of both BGL and Thunderbird datasets is small. When we use 80\% logs to train the model, we find only 6.7\% and 1.7\% log messages containing OOV words in the testing set of the BGL and Thunderbird dataset, respectively.

Figure \ref{fig:templates_with_oov} shows the percentages of log templates (produced by Drain \cite{he2017drain}) that contains OOV words on BGL and Thunderbird datasets, as the percentage of training data increases from 20\% to 80\%.
We observe that all testing sets have log templates with OOV words, even when trained with 80\% of the data. The proportion of log templates containing OOV words on the BGL dataset is always more than 80\%, no matter how much data is used for training.
The percentage of templates containing OOV words on the Thunderbird dataset decreases with the growth of training data, but still, more than 60\% when 80\% of data is trained.

The results show that a small number of log messages containing OOV words can produce many unseen log events in the testing set. There are three main reasons for this finding:
\begin{itemize}
    \item Many log events only appear during a specific period \cite{du2017deeplog}. For example, there are 842 events that only appear in the last 20\% of logs, in the 80/20 splitting.
    \item The distribution of log events is imbalanced. For example, the event \textit{``generating $*$"} appears in 1,706,751 log messages (35.95\% of the dataset), while others such as \textit{``memory manager $*$ buffer $*$"} only appear less than 100 times.
    \item OOV words can cause log parsing errors and lead to many extra log events. These extra log events usually appear a few times but still make up a majority of log templates. For example, 1,165 log events only appear once in the BGL dataset.
\end{itemize}
 
Our finding indicates that anomaly detection methods based only on log events could lead to many inaccurate detection results.
For example, SVM \cite{liang2007failure} and LR \cite{chen2004failure}, which transform log sequences into log count vectors, cannot take new log events as input because the dimension of log count vectors is fixed (i.e., the number of original log events). Moreover, DeepLog \cite{du2017deeplog}, using the indexes of log templates to predict the next log event, considers all new log events as anomalies because they cannot be predicted by the model.

\subsection{Log Parsing Errors Introduced by Semantic Misunderstanding}
\label{sec:semantic_misunderstanding}
We identify two main cases of parsing errors that are introduced by semantic misunderstanding:
\begin{itemize}
    \item Case 1: Misidentifying parameters as keywords.
    \item Case 2: Misidentifying keywords as parameters.
\end{itemize}

\begin{figure}[h]
    \vspace{-1mm}
    \centering
    \includegraphics[width=.9\linewidth]{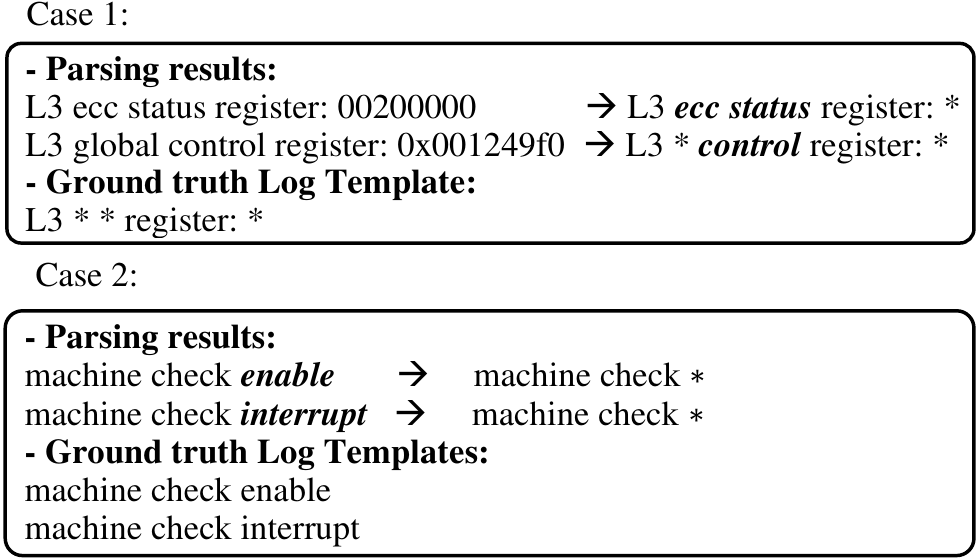}
    \caption{Examples of Log Parsing Error (Drain)}
    \label{fig:examples_log_parsing_error}
    \vspace{-1mm}
\end{figure}

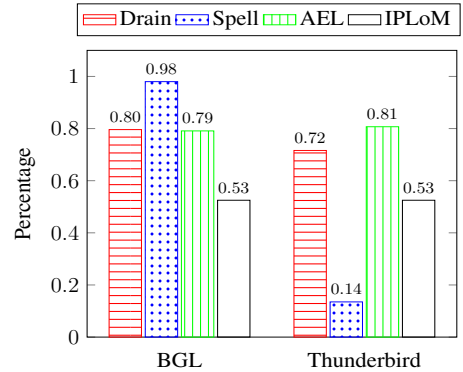
\begin{figure}[h]
\vspace{-1mm}
\centering
\resizebox{0.7\columnwidth}{.67\height}{
\begin{tikzpicture}
    \large
    \centering
    \begin{axis}[
        ybar,
        symbolic x coords={BGL, Thunderbird},
        xtick=data,
        ymin=0,
        ymax=1.1,
        ylabel={Percentage},
        major x tick style = transparent,
        bar width=0.6cm,
        enlarge x limits=0.5,
        nodes near coords={\pgfmathprintnumber[fixed zerofill,precision=2]{\pgfplotspointmeta}},
        every node near coord/.append style={
            font=\small,
            color=black
        },
        legend style={
            at={(1.01,1.05)},
            anchor=south east,
            legend columns=4,
        }
    ]
    \addplot [
        red,
        pattern=horizontal lines,
        pattern color=red,
        area legend
    ] coordinates {(BGL,0.796) (Thunderbird, 0.716)};
    \addplot [
        blue,
        pattern=dots,
        pattern color=blue,
        area legend
    ] coordinates {(BGL,0.98) (Thunderbird, 0.135)};
    \addplot [
        green,
        pattern=vertical lines,
        pattern color=green,
        area legend
    ] coordinates {(BGL,0.791) (Thunderbird, 0.807)};
    \addplot [
        black,
        area legend
    ] coordinates {(BGL,0.525) (Thunderbird, 0.525)};
    \legend{Drain, Spell, AEL, IPLoM}
    \end{axis}
\end{tikzpicture}
}
\caption{Percentages of extra log events produced by four log parsers}
\label{fig:percentage_extra_event}
\end{figure}

For Case 1, the parameters in log messages are misidentified as keywords and included in the log templates produced by the log parsers, thus leading to many extra log events. We compare the parsed template of each log message with the ground truth of the BGL dataset. If a template contains more keywords than the ground truth, it is considered wrongly parsed and an extra log event. The two log messages in Case 1 in Figure \ref{fig:examples_log_parsing_error} only refer to one log template but are parsed into two different log templates.
Figure \ref{fig:percentage_extra_event} shows the percentages of extra log events produced by the four log parsers on two datasets.
For example, there are about 80\% extra log events on the BGL dataset and 72\% extra log events on the Thunderbird dataset using Drain.

\begin{figure}[h]
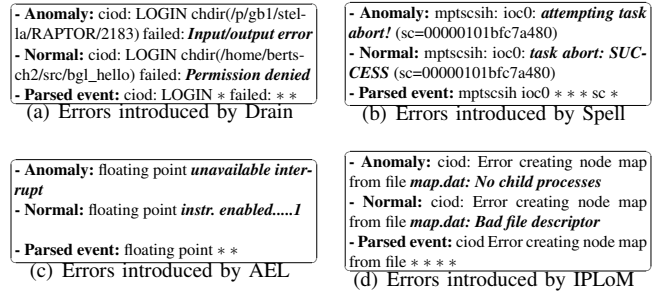

    \vspace{-1mm}
    \centering
    \subfigure[Errors introduced by Drain]{
        \centering
        \resizebox{0.46\linewidth}{!}{
            \begin{minipage}[]{\linewidth}
                \Large
                \centering
                \cornersize{.1}
                \ovalbox{\begin{minipage}{1.0\linewidth}
                
                \textbf{- Anomaly:} ciod: LOGIN chdir(/p/gb1/stel-\\la/RAPTOR/2183) failed: \textbf{\textit{Input/output error}}
                
                \textbf{- Normal:} ciod: LOGIN chdir(/home/berts-\\ch2/src/bgl\_hello) failed: \textbf{\textit{Permission denied}}
                
                \textbf{- Parsed event:} ciod: LOGIN $*$ failed: $*$ $*$
                \end{minipage}}
            \end{minipage}
        }
        \label{fig:parsing_error_drain}
    }
    \subfigure[Errors introduced by Spell]{
        \centering
        \resizebox{0.46\linewidth}{!}{
            \begin{minipage}[]{\linewidth}
                \Large
                \centering
                \cornersize{.1}
                \ovalbox{\begin{minipage}{1.0\linewidth}
                
                \textbf{- Anomaly:} mptscsih: ioc0: \textit{\textbf{attempting task abort!}} (sc=00000101bfc7a480)

                \textbf{- Normal:} mptscsih: ioc0: \textit{\textbf{task abort: SUCCESS}} (sc=00000101bfc7a480)
                
                \textbf{- Parsed event:} mptscsih ioc0 $*$ $*$ $*$ sc $*$
                \end{minipage}}
            \end{minipage}
        }
        \label{fig:parsing_error_spell}
    }
    \subfigure[Errors introduced by AEL]{
        \centering
        \resizebox{0.46\linewidth}{!}{
            \begin{minipage}[]{\linewidth}
                \Large
                \centering
                \cornersize{.1}
                \ovalbox{\begin{minipage}{1.0\linewidth}
                
                \textbf{- Anomaly:} floating point \textbf{\textit{unavailable interrupt}}
                
                \textbf{- Normal:} floating point \textbf{\textit{instr. enabled.....1}}\\
                
                \textbf{- Parsed event:} floating point $*$ $*$
                \end{minipage}}
            \end{minipage}
        }
        \label{fig:parsing_error_ael}
    }
    \subfigure[Errors introduced by IPLoM]{
        \centering
        \resizebox{0.46\linewidth}{!}{
            \begin{minipage}[]{\linewidth}
                \Large
                \centering
                \cornersize{.1}
                \ovalbox{\begin{minipage}{1.0\linewidth}
                
                \textbf{- Anomaly:} ciod: Error creating node map from file \textit{\textbf{map.dat: No child processes}}
                
                \textbf{- Normal:} ciod: Error creating node map from file \textit{\textbf{map.dat: Bad file descriptor}}
                
                \textbf{- Parsed event:} ciod Error creating node map from file $*$ $*$ $*$ $*$
                \end{minipage}}
            \end{minipage}
        }
        \label{fig:parsing_error_iplom}
    }
    \caption{{Examples of Valuable Information Removed by Log Parser}}
    \label{fig:examples_semantic_loss}
    \vspace{-1mm}
\end{figure}

For Case 2, some essential keywords in log messages could be removed after log parsing, 
resulting in different log messages being parsed into one log event. Figure \ref{fig:examples_log_parsing_error} shows an example of this case. Two different log messages are parsed into the same log event \textit{``machine check $*$"}. However, one indicates a normal behavior (i.e. \textit{``machine check enable"}), while the other indicates a system anomaly (i.e., \textit{``machine check interrupt"}). The errors of this type make the detection model difficult to distinguish between normal or abnormal logs based only on log events.
Figure \ref{fig:examples_semantic_loss} shows examples of Case 2 parsing errors introduced by the four log parsers. Each example shows one normal log and one abnormal log which are parsed into the same log event.
Valuable information such as the reason for login failure (i.e., Figure \ref{fig:parsing_error_drain}) is missing from log events and leads to many wrong detection results. 


Table \ref{tab:drain-parsed-error} shows examples of wrongly parsed log templates by the Drain parser \cite{he2017drain}. For instance, on the BGL dataset, there are 6,541 log messages that have the template \textit{``floating point $*$ $*$"}. However, only log messages in the form of \textit{``floating point unavailable interrupt"} are labeled as anomalies, while others (such as \textit{``floating point instr. enabled"} or \textit{``floating point alignment exceptions"}) indicate normal system behavior.
Similarly, Drain also produces 899 log messages that have the log templates indicating both normal and abnormal states on the Thunderbird dataset.

We also observe similar results for other log parsers. On BGL, the numbers of misidentified log messages produced by Spell, AEL, and IPLoM are 58,228, 20,154, and 31,298, respectively.
On Thunderbird, the numbers of misidentified log messages produced by Spell, AEL, and IPLoM are 3,851, 1,463, and 5,687, respectively.


\begin{table}[h]
\vspace{-1.5mm}
\caption{Examples of log parsing errors introduced by Drain} \label{tab:drain-parsed-error}
\resizebox{\columnwidth}{!}{
    \begin{tabular}{|c|c|l|l|}
    \hline
    \multicolumn{1}{|l|}{} & {\textbf{\begin{tabular}[c]{@{}c@{}}\#Anom.\end{tabular}}} & \textbf{Log Template} & \textbf{Occu.} \\ \hline
    \multirow{4}{*}{BGL} & \multirow{4}{*}{348,460} & floating point $*$ $*$ & 6,541 \\ \cline{3-4} 
     &  & machine check $*$ & 6,594 \\ \cline{3-4} 
     &  & \begin{tabular}[c]{@{}l@{}}ciod: Error creating node map from file $*$ $*$ $*$ $*$\end{tabular} & 2,952 \\ \cline{3-4} 
     &  & ciod: LOGIN $*$ failed: $*$ $*$ & 1,289 \\ \hline
    \multirow{2}{*}{Thunderbird} & \multirow{2}{*}{4,934} & \begin{tabular}[c]{@{}l@{}}mptscsih: ioc0: attempting $*$ $*$ $*$\end{tabular} & 771 \\ \cline{3-4}
     &  & \begin{tabular}[c]{@{}l@{}}EXT3-fs error (device $*$ $*$ $*$ $*$ $*$ aborted)\end{tabular} & 121 \\ \cline{3-4} 
     &  & \begin{tabular}[c]{@{}l@{}}Out of Memory: Killed process $*$ $*$\end{tabular} & 7 \\ \hline
    \end{tabular}
}
\begin{tablenotes}
      \footnotesize
      \item Note: $\#Anom.$ denotes the number of anomalies. $Occu.$ denotes the number of occurrences of the log template.
\end{tablenotes}
\vspace{-1.5mm}
\end{table}

\subsection{The Impact of Log Parsing Errors on Anomaly Detection}
\label{sec:impract_of_logparsing_errors}

The existing approaches share a common process: they all utilize a log parser to parse the log messages into log events (i.e., the templates of log messages), construct log sequences, and then build unsupervised or supervised machine learning models to detect anomalies. The existing approaches can be adversely affected by the log parsing errors introduced by the OOV words and semantic misunderstanding.

In this section, we evaluate the impact of log parsing errors on two representative anomaly detection methods SVM-based method \cite{liang2007failure}, and LogRobust \cite{zhang2019robust}. SVM represents those ML-based approaches that use the log count vectors as input. LogRobust represents recent DL-based approaches that utilize the semantic vectors of log templates as input.
They both use a log parser to generate a set of log events. SVM-based method \cite{liang2007failure} represents log sequences as log count vectors 
and then constructs a hyperplane to separate normal and abnormal samples in a high-dimension space. LogRobust \cite{zhang2019robust} incorporates a pre-trained Word2vec model \cite{joulin2016fasttext} to learn the semantic vectors of log templates instead of counting log events' occurrences. Using the Word2vec model allows LogRobust to discover the semantic relationship between log events and handle the instability of log data.

\begin{figure}[h]
    \centering
    \subfigure[SVM on BGL] {
        \includegraphics[scale=0.47]{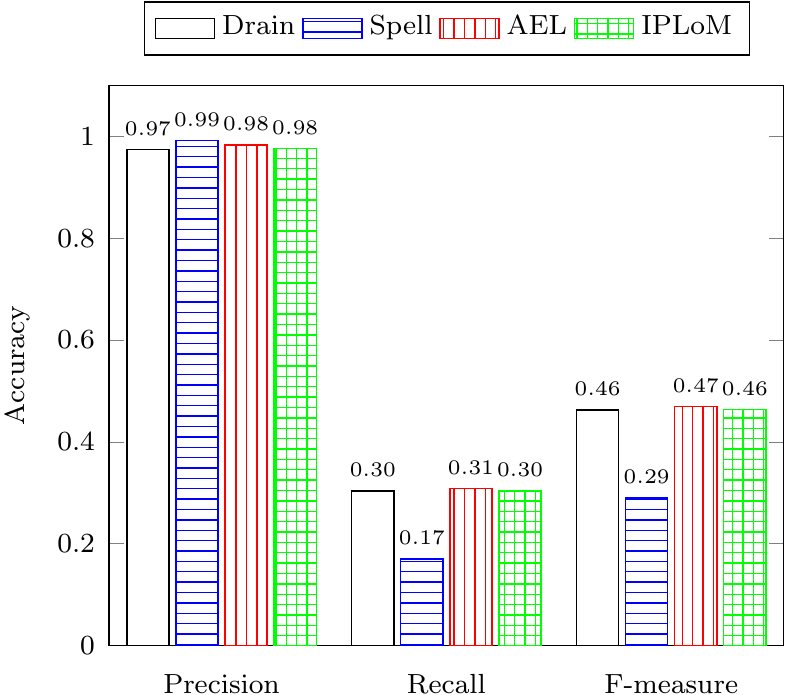}
        \label{fig:bgl_svm_accuracy}
    }
    \subfigure[LogRobust on BGL] {
        \includegraphics[scale=0.47]{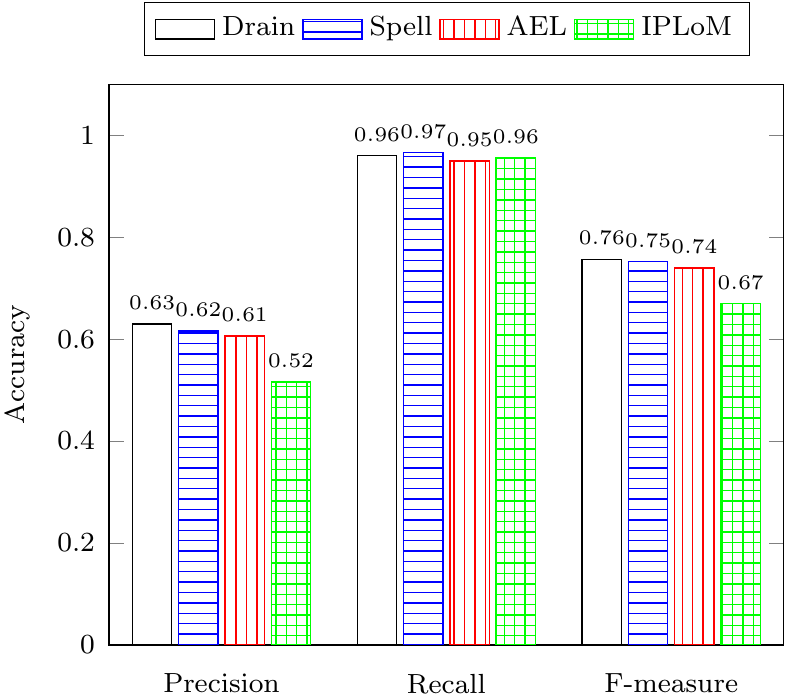}
        \label{fig:bgl_logrobust_accuracy}
    }
    \subfigure[SVM on Thunderbird]{
    \includegraphics[scale=0.47]{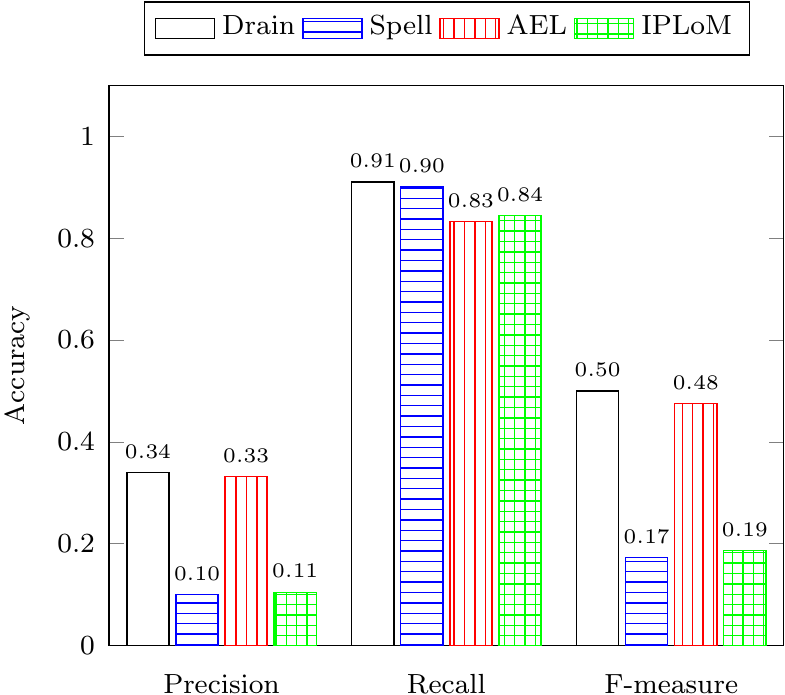}
    \label{fig:thunderbird_SVM_accuracy}
    }
    \subfigure[LogRobust on Thunderbird]{
    \includegraphics[scale=0.47]{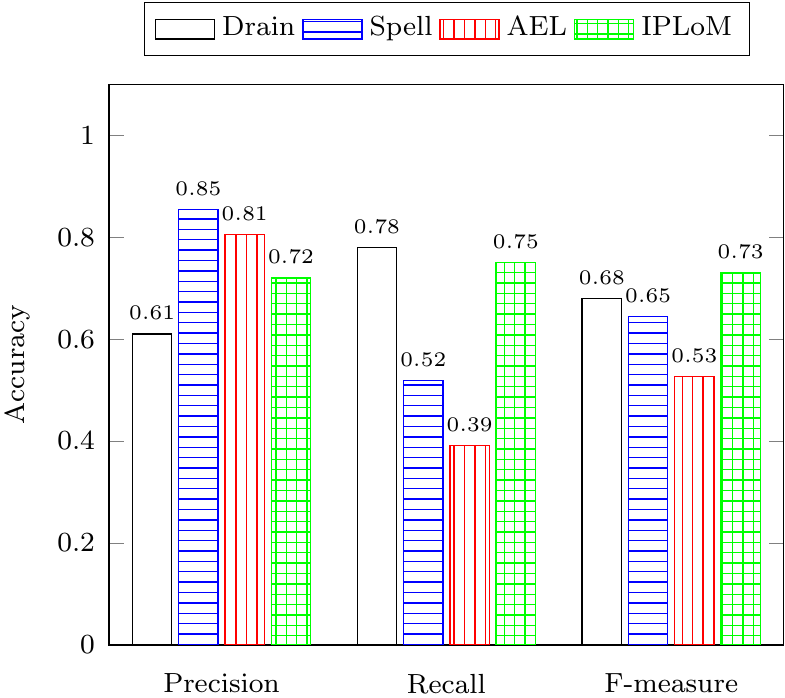}
    \label{fig:thunderbird_logrobust_accuracy}
    }
    \caption{Results of anomaly detection with different log parsers on BGL and Thunderbird datasets}
    \label{fig:different_parsing_result}
\end{figure}

Figure \ref{fig:different_parsing_result} shows the results of the SVM-based method and LogRobust with four different log parsers (i.e., Drain \cite{he2017drain}, Spell \cite{du2016spell}, AEL \cite{jiang2008abstracting}, and IPLoM \cite{makanju2009clustering}). We observe that the performance of current anomaly detection methods 
is affected by the accuracy of log parsing, and different log parsers could lead to different results.
SVM achieves better results when using Drain \cite{he2017drain} and AEL \cite{jiang2008abstracting} since these parsers produce a smaller amount of inaccurate log events, as discussed in Section \ref{sec:semantic_misunderstanding}. Figure \ref{fig:bgl_svm_accuracy} and Figure \ref{fig:thunderbird_SVM_accuracy} show that SVM with Drain achieves an F1-score of 0.46 and 0.50 on BGL and Thunderbird datasets, respectively. While SVM with Spell only produces F1-scores of 0.29 on the BGL dataset and 0.17 on the Thunderbird dataset.
LogRobust can achieve better results than SVM since LogRobust can identify unstable log events with similar semantic meaning through semantic vectorization. Still, LogRobust suffers from the log parsing errors caused by semantic misunderstanding (see Section \ref{sec:semantic_misunderstanding}) and achieves F1-scores of less than 0.8 on both datasets.

\begin{figure}[h]
    \vspace{-1mm}
    \centering
    \subfigure[Accuracy of SVM] {
        \large
        \resizebox{0.46\columnwidth}{!}{
        \begin{tikzpicture}
            \large
            \begin{axis}[
                ybar,
                symbolic x coords={Drain, Spell, AEL, IPLoM},
                xtick=data,
                ymin=0,
                ymax=1.1,
                ylabel={F-measure},
                major x tick style = transparent,
                bar width=0.6cm,
                enlarge x limits=0.15,
                nodes near coords={\pgfmathprintnumber[fixed zerofill,precision=2]{\pgfplotspointmeta}},
                every node near coord/.append style={
                    font=\small,
                    color=black
                },
                legend style={
                    at={(0.75,1.05)},
                    anchor=south east,
                    legend columns=2,
                },
            ]
            \addplot [
                red,
                pattern=horizontal lines,
                pattern color=red,
                area legend
            ] coordinates {(Drain,0.46) (Spell, 0.29) (AEL, 0.47) (IPLoM, 0.46)};
            \addplot [
                green,
                pattern=dots,
                pattern color=green,
                area legend
            ] coordinates {(Drain,0.54) (Spell, 0.54) (AEL, 0.54) (IPLoM, 0.54)};
            \legend{Before, After}
        \end{axis}
        \end{tikzpicture}
        }
        \label{fig:result_after_fixing_svm}
    }
    \subfigure[Accuracy of LogRobust]{
        \large
        \resizebox{0.46\columnwidth}{!}{
            \begin{tikzpicture}
            \large
            \begin{axis}[
                ybar,
                symbolic x coords={Drain, Spell, AEL, IPLoM},
                xtick=data,
                ymin=0,
                ymax=1.1,
                ylabel={F-measure},
                major x tick style = transparent,
                bar width=0.6cm,
                enlarge x limits=0.15,
                nodes near coords={\pgfmathprintnumber[fixed zerofill,precision=2]{\pgfplotspointmeta}},
                every node near coord/.append style={
                    font=\small,
                    color=black
                },
                legend style={
                    at={(0.75,1.05)},
                    anchor=south east,
                    legend columns=2,
                },
            ]
            \addplot [
                red,
                pattern=horizontal lines,
                pattern color=red,
                area legend
            ] coordinates {(Drain,0.76) (Spell, 0.75) (AEL, 0.74) (IPLoM, 0.67)};
            \addplot [
                green,
                pattern=dots,
                pattern color=green,
                area legend
            ] coordinates {(Drain,0.91) (Spell, 0.91) (AEL, 0.91) (IPLoM, 0.91)};
            \legend{Before, After}
        \end{axis}
        \end{tikzpicture}
        }
        \label{fig:result_after_fixing_logrobust}
    }
\caption{The accuracy of anomaly detection before and after fixing the log parsing errors on the BGL dataset}
\label{fig:result_after_fixing}
\vspace{-1mm}
\end{figure}
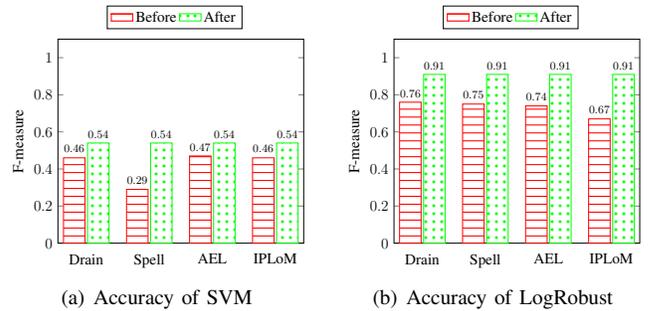

Next, we manually fix the errors produced by log parsing methods on the BGL dataset, then apply SVM and LogRobust to confirm whether or not the accuracy of anomaly detection methods is improved if log parsing performs more accurately. We leverage the ground truth log templates for the BGL dataset from \cite{logpai}.
For each wrongly parsed log message, we match it with the most similar log template in the ground truth. 
After the fixing process, the outputs produced by the log parser are actually the ground truth. Also, the outputs of different log parsers are the same after fixing (as they are all the same as the ground truth).
Figure \ref{fig:result_after_fixing} shows the accuracy (measured in terms of F-measure) of each individual parser before and after fixing the log parsing errors. 
We can see that both SVM and LogRobust perform better when the log parsing errors are fixed (on average, 25\% improvement for LogRobust and 29\% improvement for SVM).

Overall, the results show that log parsing accuracy affects the performance of anomaly detection. 
Existing log parsing methods cannot handle well the OOV words in new logs, thus losing semantic information while detecting anomalies. Furthermore, current log parsing methods 
could produce errors due to semantic misunderstanding. Therefore, existing anomaly detection methods that leverage log events
are unable to achieve satisfying results due to the imperfections of log parsing methods.
\section{NeuralLog: Log-based Anomaly Detection Without Log Parsing}
\label{sec:approach}
\begin{figure}[h]
\vspace{-1mm}
  \centering
  \includegraphics[width=0.9\linewidth]{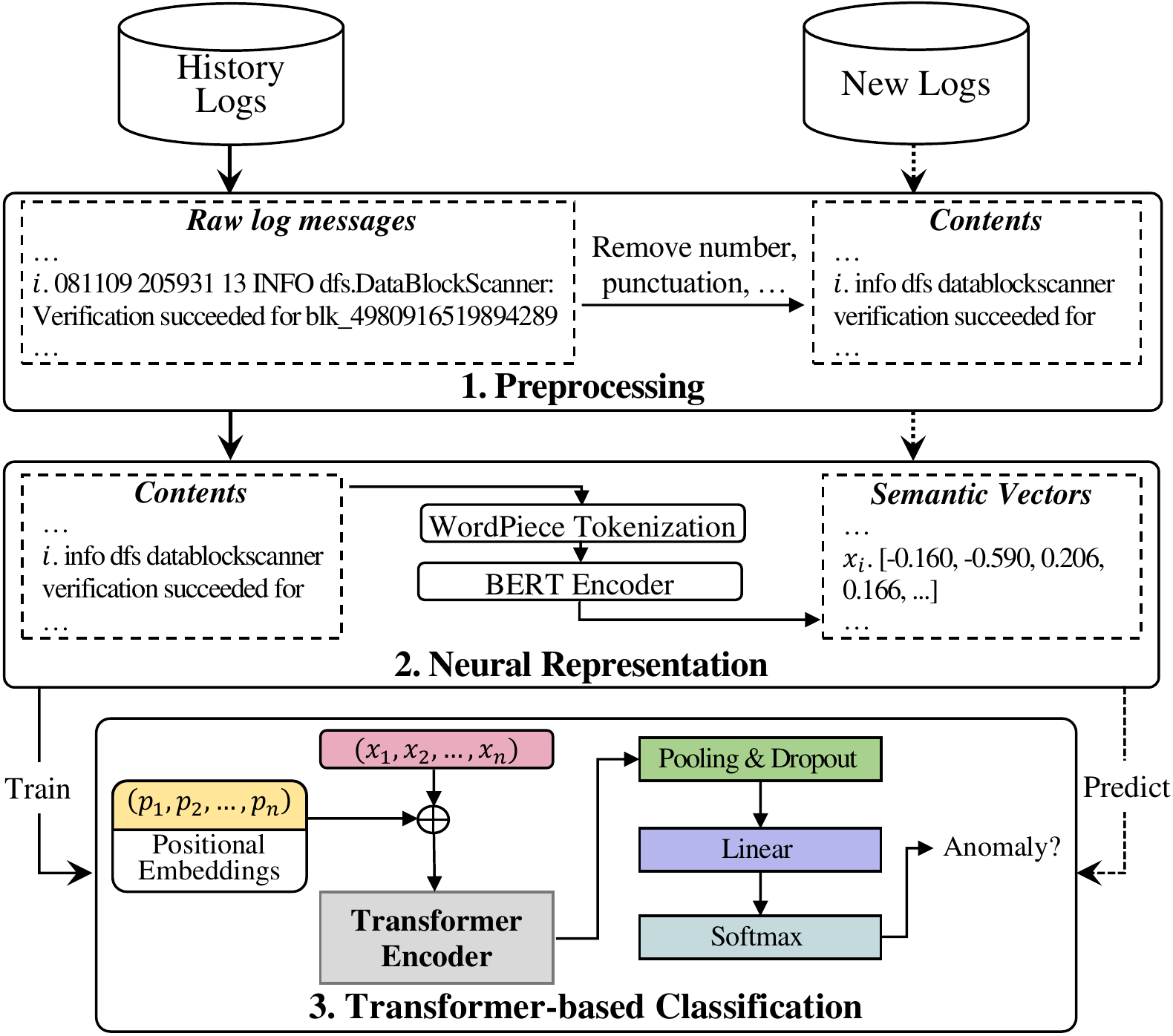}
  \caption{An Overview of \tool}
  \label{fig:overview}
\vspace{-1mm}
\end{figure}

To overcome the limitation of existing approaches, 
we propose NeuralLog, a new log-based anomaly detection approach that directly uses raw log messages to detect anomalies. The overview of the proposed approach is shown in Figure \ref{fig:overview}. Overall, \tool consists of three steps: preprocessing (Section \ref{sec:preprocessing}), neural representation (Section \ref{sec:representation}), and transformer-based classification (Section \ref{sec:classification}). The first step is log preprocessing. After that, each log message is encoded into a semantic vector by using BERT. In this way, our approach can prevent the loss of valuable information from log messages. Finally, we leverage the Transformer \cite{vaswani2017attention} model to detect the anomalies. 

\subsection{Preprocessing}
\label{sec:preprocessing}
Preprocessing log data is the first step for building our model. 
In this step, we first tokenize a log message into a set of word tokens. We use common delimiters in the logging system (i.e., white space, colon, comma, etc.) to split a log message. Then, every capital letter is converted to a lower letter, and we remove all non-character tokens from the word set. These non-characters contain operators, punctuation marks, and number digits. This type of tokens is removed since it usually represents variables in the log message and is not informative.
As an example, the raw log message
\textit{``081109 205931 13 INFO dfs.DataBlockScanner: Verification succeeded for blk\_-4980916519894289629"}
is first split into a set of words based on common delimiters. Then non-character tokens are excluded from the set. Finally, a set of words {\textit{\{info, dfs, datablockscanner, verification, succeeded}\}} is obtained.





\subsection{Neural Representation}
\label{sec:representation}
Each log message records a system event with its header and message content. The message header contains fields determined by the logging framework, such as component and verbosity level. The message content written by developers reflects a specific state of the system. Existing methods usually analyze only message content and remove other information.
In this paper, \tool uses all textual information such as verbosity, component, and content to extract the semantic meaning of log messages. In order to reserve semantic information and capture relationships among existing and new log messages, the representation phase tries to represent log messages in the vector format.


\subsubsection{Subword Tokenization}


Tokenization can be considered as the first step to handle OOV words. In our work, we adopt the WordPiece tokenization \cite{schuster2012japanese, wu2016google}, which is widely used in many recent language modeling studies \cite{devlin2018bert, sanh2019distilbert, clark2020electra}.

WordPiece includes all the characters and symbols into its base vocabulary first.
Instead of relying on the frequency of the pairs, WordPiece chooses the one that maximizes the training data's likelihood. It trains a language model starting from the base vocabulary and picks the pair with the highest likelihood.
This pair is added to the vocabulary, and the language model is again trained on the new vocabulary. These steps are repeated until the desired vocabulary is reached.
For example, the rare word \textit{``datablockscanner"} is split into more frequent subwords: \{\textit{``data"}, \textit{``block"}, \textit{``scan"}, \textit{``ner"}\}.
In this way, the number of OOV words is reduced and their meanings are captured.

The reason we choose WordPiece is that it can effectively handle the OOV words and reduce the vocabulary size. Compared with other tokenization (chunking) approaches, WordPiece is more effective.
For example, space/stemming/camel case based tokenization strategies can lead to many OOV words and a big vocabulary \cite{karampatsis2020big}.

\subsubsection{Log Message Representation}
After preprocessing and tokenization, \tool transforms each log message into a set of words and subwords.
Conventionally, words of log content are further transformed into vectors by using Word2Vec \cite{le2014distributed}, then the representation vector of each sentence would be calculated based on the word vectors. However, Word2Vec produces the same embedding for the same word. In many cases, a word can have different meanings based on its position and context. 
BERT \cite{devlin2018bert} is a recent deep learning representation model that has been pre-trained on a huge natural language corpus. 
In our work, we employ the feature extraction function of pre-trained BERT to obtain the semantic meaning of log messages.




More specifically, after tokenizing, the set of words and subwords is passed to the BERT model and encoded into a vector representation with a fixed dimension. 
\tool utilizes the BERT base model \cite{google2021bertpretrained} that contains 12-layers of transformer encoder and 768-hidden units of each transformer. Each layer generates embeddings for each subword in a log message. We use the word embeddings generated by the last encoder layer of BERT in our work. Then, the embedding of a log message is calculated as the average of its corresponding word embeddings.
As any word that does not occur in the vocabulary (i.e., OOV words) is broken down into subwords, BERT can learn the representation vector of those OOV words based on the meaning of subword collections. Besides, the positional embedding layer allows BERT to capture the representation of a word based on its context in a log message.
BERT also contains self-attention mechanisms that can effectively measure the importance of each word in a sentence. 

\subsection{Transformer-based Classification}
\label{sec:classification}

To better understand the semantics of logs, we adopt the transformer model \cite{vaswani2017attention}, which has been introduced to overcome the limitations of RNN-based models. 
Taking the semantic vectors of log messages as input (i.e. $\mathbf{X} = \{x_1, x_2, \dots, x_n\}$), we use a transformer encoder-based model for anomaly detection. In this section, we briefly describe the proposed transformer-based classification model, which contains Positional Encoding and Transformer Encoder.

\paragraph{Positional Encoding}
\label{sec:positional_encoding}
The order of a log sequence conveys important information for the anomaly detection task. BERT encoder represents a log message into a fixed-dimensional
vector where log messages with similar meanings are closer to each other. However, those vectors do not contain the relative position information of log messages in a log sequence.
Therefore, a sinusoidal encoder is applied to generate
an embedding $p_i$ using $sin$ and $cos$ functions for each position $i$ in the log sequence $\mathbf{X}$ \cite{vaswani2017attention}.
Then, $p_i$ is added to the semantic vector $x_i$ at position $i$, and $x_i + p_i$ will be used to feed the transformer-based model (see Figure \ref{fig:overview} (Step 3)).
In this way, the model can learn the relative position information of each log message in the sequence and can distinguish log messages at different positions.

\paragraph{Transformer Encoder}

This model is based on the transformer architecture \cite{vaswani2017attention}, which contains self-attention layers followed by position-wise feed-forward layers. 
Given an input $\mathbf{X} = \{x_1, x_2, \dots, x_n\}$, the positional embeddings are added before it enters into the transformer. In the transformer module, multi-head attention layers calculate the attention score matrices for each log message with different attention patterns.
The attention score is calculated by training the query and key matrices of the attention layers. Different attention patterns are obtained with multi-head self-attention layers, which enable the model to consider which attention score is significant. 
The inter-layer features are connected into a feed-forward network, which contains two fully connected layers in order to reach the combination of different attention scores. Then, the output of the transformer model is fed into the pooling, dropout, and a fully connected layer. The class probabilities, which identify normal/abnormal log sequences, are calculated using the softmax classifier. The architecture of the classification model is shown in Figure \ref{fig:overview} (Step 3).


\subsection{Anomaly Detection}
\label{sec:anomaly_detection}
Following the above steps, we can train a
transformer-based model for log-based anomaly detection.
When a set of new log messages arrives, \tool firstly conducts preprocessing. Then it transforms the new log messages into semantic vectors. The log sequence,
represented as a list of semantic vectors, is fed into the trained model. Finally, the transformer-based model can predict whether this log sequence is anomalous or not.
\vspace{-1mm}
\section{Evaluation}
\label{sec:evaluation}
\subsection{Experimental Design}

\subsubsection{Research Questions}

In this section, we evaluate our approach by answering the following research questions (RQs):


RQ1: How effective is \tool in log-based anomaly detection?

RQ2: How effective is \tool in understanding the semantic meaning of log data?



RQ3: How effective is \tool under different settings?

\subsubsection{Datasets}
In this paper, we evaluate \tool on four public datasets \cite{loghub}, namely HDFS, Blue Gene/L, Thunderbird, and Spirit. HDFS dataset \cite{he2020loghub, xu2009detecting} contains 11,175,629 log messages collected from a Hadoop Distributed File System on the Amazon EC2 platform. Each session identified by block ID in the HDFS dataset is labeled as normal or abnormal. BGL dataset \cite{oliner2007supercomputers, he2020loghub} contains 4,747,963 log messages collected from the Blue Gene/L supercomputer at Lawrence Livermore National Labs. 
Thunderbird and Spirit datasets \cite{oliner2007supercomputers} were collected from two real-world supercomputers at Sandia National Labs. Each log message in these datasets was manually labeled as anomalous or not. 
In this experiment, we leverage 10 million continuous log messages from the Thunderbird dataset, and 1GB log messages
from the Spirit dataset, which were also used
in prior work \cite{yao2020study}.
The details of the datasets are shown in Table \ref{tab:dataset_detail}.

\begin{table}[h]
    \vspace{-1.5mm}
    \caption{The details of log datasets}
    \label{tab:dataset_detail}
    \resizebox{\columnwidth}{!}{%
    \setlength{\tabcolsep}{4pt}
    \begin{tabular}{@{}ccccc@{}}
    \toprule
    \multicolumn{1}{c}{} & \multicolumn{1}{c}{\textbf{Category}} & \multicolumn{1}{c}{\textbf{Size}} & \multicolumn{1}{c}{\textbf{\#Messages}} & \multicolumn{1}{c}{\textbf{\#Anomalies}} \\ \midrule
    HDFS         & Distributed system & 1.5 G & 11,175,629 & 16,838 \\
    Blue Gene /L & Supercomputer      & 743 M & 4,747,963  & 348,460      \\
    Thunderbird  & Supercomputer      & 1.4 G & 10,000,000 & 4,934     \\
    Spirit       & Supercomputer      & 1.0 G & 7,983,345  & 768,142    \\ \bottomrule
    \end{tabular}
    }
    \vspace{-1mm}
\end{table}

\subsubsection{Implementation and Environment}
In our experiments, \tool has one layer of the transformer encoder.
The number of attention heads is 12, and the size of the feed-forward network that takes the output of the multi-head self-attention mechanism is 2048. 
The Transformer-based model of \tool is trained using AdamW optimizer \cite{loshchilov2017decoupled} with the initial learning rate of $3e-4$. We set the mini-batch size and the dropout rate to 64 and 0.1, respectively. We use the cross-entropy as the loss function. We train the Transformer-based model for a maximum of 20 epochs and perform early stopping for five consecutive iterations.

We implement \tool with Python 3.6 and Keras toolbox and conduct experiments on a server with Windows Server 2012 R2, Intel Xeon E5-2609 CPU, 128GB RAM, and an NVIDIA Tesla K40c.

\subsubsection{Evaluation Metrics}
To measure the effectiveness of \tool in anomaly detection, we use the Precision, Recall, and F1-Score metrics. We calculate these metrics as follows:

\begin{itemize}
    \item \textit{Precision:} the percentage of correctly detected abnormal log sequences amongst all detected abnormal log sequences by the model. $Precision = \frac{TP}{TP + FP}$.
    \item \textit{Recall:} the percentage of log sequences that are correctly identified as anomalies over all real anomalies.\\ $Recall = \frac{TP}{TP + FN}$.
    \item \textit{F1-Score:} the harmonic mean of \textit{Precision} and \textit{Recall}.\\ $F1-score = \frac{2 * Precision * Recall}{Precision + Recall}$
\end{itemize}
TP (True Positive) is the number of abnormal log sequences the are correctly detected by the model. FP (False Positive) is the number of normal log sequences that are wrongly identified as anomalies. FN (False Negative) is the number of abnormal log sequences that are not detected by the model.

\subsection{RQ1: How effective is NeuralLog?}

This RQ evaluates whether or not \tool can work effectively on public log datasets. For the HDFS dataset, we construct log sequences by correlating log messages with the same block ID, as the data is labeled by blocks. Then, we randomly select 80\% of log sequences for training, and the rest of the dataset is used for testing. For BGL, Thunderbird, and Spirit datasets, we first sort the log messages by time. Then, we leverage the first 80\% (according to the timestamps of logs) log messages as the training set and the rest 20\% as the testing set. This design ensures that the testing data contains new log messages previously unseen in the training set. Following the previous work \cite{li2020_swisslog, meng2019loganomaly}, we apply a sliding window with a length of 20 messages and a step size of 1 message to construct log sequences.

\begin{table}[h]
\vspace{-1.5mm}
\caption{Results of different methods on public datasets}
\label{tab:RQ1_results}
\resizebox{\columnwidth}{!}{%
\setlength{\tabcolsep}{4pt}
\renewcommand{\arraystretch}{0.9}
\begin{tabular}{@{}clccccccc@{}}
\toprule
\textbf{Dataset} & \textbf{} & \textbf{LR}  & \textbf{SVM} & \textbf{IM} & \textbf{LogRobust} & \textbf{Log2Vec} & \textbf{NeuralLog} \\ \midrule
                 & P       & 0.99         & 0.99         & 1.00         &0.98               & 0.94             & 0.96               \\
HDFS             & R          & 0.92         & 0.94         & 0.88         & 1.00               & 0.94             & 1.00               \\
                 & F1        & 0.96         & 0.96         & 0.94         & \textbf{0.99}      & 0.94             & 0.98               \\ \midrule
                 & P       & 0.13         & 0.97         & 0.13         & 0.62               & 0.80             & 0.98               \\
BGL              & R          & 0.93         & 0.30         & 0.30        & 0.96               & 0.98             & 0.98               \\
                 & F1        & 0.23         & 0.46         & 0.18        & 0.75               & 0.88             & \textbf{0.98}      \\ \midrule
\multirow{3}{*}{\begin{tabular}[c]{@{}c@{}}Thunder-\\ bird\end{tabular}}                 & P       & 0.46         & 0.34         & -          & 0.61               & 0.74             & 0.93               \\
                 & R          & 0.91         & 0.91         & -         & 0.78               & 0.94             & 1.00               \\
                 & F1        & 0.61         & 0.50         & -           & 0.68               & 0.84             & \textbf{0.96}      \\ \midrule
                 & P       & 0.89         & 0.88         & -           & 0.97               & 0.91             & 0.98               \\
Spirit           & R          & 0.96         & 1.00         & -      & 0.94               & 0.96             & 0.96               \\
                 & F1        & 0.92         & 0.93         & -           & 0.95               & 0.95             & \textbf{0.97}      \\ \bottomrule
\end{tabular}
}
\begin{tablenotes}
      \footnotesize
      \item '-' denotes timeout (30 hours), P denotes Precision, R denotes Recall, and F1 is the F1-score.
\end{tablenotes}
\vspace{-1mm}
\end{table}

We compare the results of \tool and five existing approaches, including Support Vector Machine-based approach \textbf{(SVM)} \cite{chen2004failure}, Logistic Regression-based approach \textbf{(LR)} \cite{bodik2010fingerprinting}, Invariant Mining \textbf{(IM)} \cite{lou2010mining}, \textbf{LogRobust} \cite{zhang2019robust}, and \textbf{Log2Vec} \cite{meng2020semantic}. 
Traditional approaches, such as SVM, LR, and IM, transform the log sequences into \textit{log count vectors}, then build unsupervised or supervised machine learning models to detect anomalies. In our work, we utilize Drain \cite{he2017drain} to generate the log events for SVM, LR, and IM.
LogRobust incorporates a pre-trained Word2vec model \cite{joulin2016fasttext} to learn the representations vector of log templates instead of counting the occurrences of log events. LogRobust then leverages an Attention-based Bi-LSTM to learn and detect anomalies. Log2Vec \cite{meng2020semantic} accurately extracts the semantic and syntax information from log messages and leverages the Deeplog \cite{du2017deeplog} model to improve the accuracy of anomaly detection.
We do not compare with DeepLog \cite{du2017deeplog} because previous studies already showed that Log2Vec outperforms DeepLog \cite{meng2020semantic}.
Note that there are some other recent state-of-the-art methods such as LogAnomaly \cite{meng2019loganomaly}. However, LogAnomaly \cite{meng2019loganomaly} has no publicly available implementation and requires operators' domain knowledge (to manually add domain-specific synonyms
and antonyms). Therefore, 
it is not experimentally compared in this paper.

The comparison results are shown in Table \ref{tab:RQ1_results}. Overall, \tool achieves the best results on BGL, Thunderbird, and Spirit datasets and comparable results on the HDFS dataset. 
It is worth noting that the recall value achieved by \tool on the HDFS dataset is 1.00, which means that \tool can identify all anomalies captured by the dataset with high precision. \tool achieves the best F1-score of 0.98 on the BGL dataset, 0.96 on the Thunderbird dataset, and 0.97 on the Spirit data.

As discussed in Section \ref{sec:log-based-anomaly-detection}, existing approaches (including SVM, Decision Tree, and LR) are heavily affected by the accuracy of log parsing. Besides, these approaches cannot capture the semantic information of log messages. Therefore, these approaches perform poorly on BGL and Thunderbird datasets when the log parsing is inaccurate. They can achieve a high F1-Score on the Spirit dataset since the parsing error rate is only 0.1\% for this dataset.

LogRobust \cite{zhang2019robust}, which encodes log templates into semantic vectors using the FastText pre-trained model \cite{joulin2016fasttext}, cannot work well on 2 out of 4 datasets. LogRobust shows a lower F1-Score of 0.75 and 0.68 on BGL and Thunderbird datasets, respectively. The main reason is that LogRobust utilizes the Drain log parser \cite{he2017drain} to obtain log templates. As aforementioned in Section \ref{sec:impract_of_logparsing_errors}, the Drain parser could inaccurately parse a noticeable number of log messages on BGL and Thunderbird datasets. 
Log2Vec \cite{meng2020semantic} transforms raw log messages into semantic vectors, thus can avoid errors from log parsers. Besides, Log2Vec also adopts MIMICK \cite{pinter2017mimicking}, an approach of inducing word embedding from character-level features to handle OOV words to improve anomaly detection performance. However, it is hard to extract contextual information from characters to form meaningful words \cite{sasaki2019subword, zhao2018generalizing, hu2019few}. Therefore, Log2Vec could not effectively handle some domain-specific words, such as technical terms or entity names \cite{sasaki2019subword, zhao2018generalizing, hu2019few}. Consequently, compared to \tool that uses subword-level feature to handle OOV words, Log2Vec achieves lower F1-scores on BGL and Thunderbird datasets (0.88 and 0.84, respectively).

We also evaluate the time efficiency of NeuralLog on the four datasets. On average, it takes \tool 14.3 minutes per dataset to encode all log messages and 5.2 minutes to train a detection model (20 epochs). The average time of the anomaly detection phase is 3.1 milliseconds per log sequence. Baseline methods that require log parsing take 102 minutes on average for preprocessing. Log2Vec spends an average of 314 minutes in the preprocessing phase. 
SVM and LR models can finish training within 0.5$\sim$0.7 minutes. LogRobust and Log2Vec can train a detection model in an average of 2.2 and 13.1 minutes, respectively. In the detection phase, it takes LogRobust 0.2 milliseconds per log sequence, and it is 26.3 milliseconds for Log2Vec.
\tool can scale to large datasets. For example, NeuralLog is able to handle the HDFS dataset, which contains 11,175,629 log messages.
It took NeuraLog 19.7 minutes for preprocessing, 7.2 minutes for training, and 0.6 minutes for testing to perform the experiment for this RQ on the HDFS dataset.

In summary, the experimental results confirm that \tool can work effectively and efficiently for log-based anomaly detection.



\subsection{RQ2: How effective is \tool in understanding the semantic meaning of log data?}
\vspace{-1mm}
In this section, we evaluate the ability of \tool to capture the semantic meaning of log messages. To this end, we examine the effectiveness of the encoding component that represents log messages as semantic vectors and the subword tokenization component that handles OOV words.

In NeuralLog, we preprocess the raw log messages and directly encode the preprocessed log messages into semantic vectors. We compare \tool with two variants:
\begin{itemize}
    \item NeuralLog-Index: the indexes of log templates, obtained by Drain \cite{he2017drain}, are simply encoded into numeric vectors and passed to the Transformer model for anomaly detection. The rest of \tool is kept the same.
    \item NeuralLog-Template: we utilize BERT to encode the log templates produced by the Drain \cite{he2017drain} into semantic vectors. We then feed these semantic vectors to the Transformer model for anomaly detection. The rest of \tool is kept the same.
\end{itemize}

Table \ref{tab:RQ2_results} shows the results of two variants of NeuralLog.
We can see that, on HDFS and Spirit dataset, these two variants can achieve high F1-scores. 
The reason is that log parsers perform well on these datasets. We find that the parsing error rate on the Spirit dataset is only 0.1\%. Besides, the HDFS system records relatively simple operations with only 29 event types, making log parsers easy to analyze. In contrast, the results of the variants on BGL and Thunderbird datasets are greatly affected by log parsing methods because they cannot precisely represent the meaning of log messages, especially when using the indexes of log templates. For example, the model using log templates' indexes and template embeddings only achieve F1-scores of 0.46 and 0.90 on the BGL dataset, which are much lower than the 0.98 F1-score achieved by \tool (which uses raw log messages).

\begin{table}[h]
\vspace{-1.5mm}
\caption{Results of different representation methods}
\label{tab:RQ2_results}
\centering
\resizebox{\columnwidth}{!}{%
\setlength{\tabcolsep}{6pt}
\renewcommand{\arraystretch}{0.9}
\begin{tabular}{@{}clcccc@{}}
\toprule
\textbf{Dataset} & \textbf{Metric} & \textbf{\begin{tabular}[c]{@{}c@{}}NeuralLog-\\Index\end{tabular}}   & \textbf{\begin{tabular}[c]{@{}c@{}}NeuralLog-\\Template\end{tabular}}   & \textbf{\tool}  \\ \midrule
                 & Precision       & 0.93                   & 0.93                      & 0.96                  \\
HDFS             & Recall          & 1.00                   & 1.00                      & 1.00                  \\
                 & F1-Score        & 0.96                   & 0.96                      & \textbf{0.98}         \\ \midrule
                 & Precision       & 0.98                   & 0.92                      & 0.98                  \\
BGL              & Recall          & 0.30                   & 0.88                      & 0.98                  \\
                 & F1-Score        & 0.46                   & 0.90                      & \textbf{0.98}         \\ \midrule
                 & Precision       & 0.58                   & 0.89                      & 0.93                  \\
Thunderbird      & Recall          & 0.98                   & 0.91                      & 1.00                  \\
                 & F1-Score        & 0.73                   & 0.90                      & \textbf{0.96}         \\ \midrule
                 & Precision       & 0.96                   & 0.93                      & 0.98                  \\
Spirit           & Recall          & 0.95                   & 0.95                      & 0.96                  \\
                 & F1-Score        & 0.95                   & 0.94                      & \textbf{0.97}         \\ \bottomrule
\end{tabular}
}
\vspace{-1mm}
\end{table}

We next evaluate whether or not \tool can effectively handle OOV words. \tool utilizes WordPiece \cite{schuster2012japanese, wu2016google} to split an OOV word into a set of subwords and then extracts the embedding of the OOV words based on its subwords.
We compare \tool with two variants:
\begin{itemize}
    \item NeuralLog-Word2Vec: We use a pre-trained Word2vec model \cite{joulin2016fasttext} to generate the embeddings of log messages. Those words that do not exist in the vocabulary are removed from log messages. Then, embedding vectors are passed to the Transformer model to detect anomalies.
    \item NeuralLog-NoWordPiece: We exclude WordPiece tokenizer from the model (see Figure \ref{fig:overview}). Log messages, after preprocessing, are directly input to the BERT model to obtain the semantic vectors. These vectors are then input to the Transformer model for anomaly detection. In this way, OOV words that do not exist in the vocabulary will be removed instead of broken down into subwords. 
\end{itemize}

The experimental results are shown in Table \ref{tab:RQ3_results_1}. \tool achieves the best performance since it utilizes WordPiece \cite{schuster2012japanese, wu2016google} to tokenize an OOV word into a set of subwords. Therefore, the meaning of an unseen word is kept by its subwords. Both variants 
achieve lower F1-scores than \tool since they rely on a fixed-size vocabulary and cannot handle the OOV words. 
For example, on the Thunderbird dataset, F1-scores achieved by NeuralLog-Word2Vec and NeuralLog-NoWordPiece are 0.80 and 0.90, respectively, which are much lower than the F1-score of 0.96 achieved by NeuralLog.

In summary, our results show that \tool can effectively represent the semantic meaning of log messages. 
Since \tool uses raw log messages (after preprocessing) for anomaly detection, the problem of inaccurate log parsing can be avoided. 
The results also show that \tool can effectively learn the meaning of OOV words. 

\begin{table}[h]
\vspace{-1.5mm}
\caption{Results of handling OOV words}
\label{tab:RQ3_results_1}
\centering
\resizebox{.95\columnwidth}{!}{%
\setlength{\tabcolsep}{6pt}
\renewcommand{\arraystretch}{0.9}
\begin{tabular}{@{}clcccc@{}}
\toprule
\textbf{Dataset} & \textbf{Metric} & \textbf{\begin{tabular}[c]{@{}c@{}}NeuralLog-\\Word2Vec\end{tabular}}      & \textbf{\begin{tabular}[c]{@{}c@{}}NeuralLog-\\NoWordPiece\end{tabular}}    & \textbf{\tool}   \\ \midrule
                 & Precision       & 0.94                   & 0.94                      & 0.96                  \\
HDFS             & Recall          & 0.93                   & 1.00                      & 1.00                  \\
                 & F1-Score        & 0.94                   & 0.97                      & \textbf{0.98}         \\ \midrule
                 & Precision       & 0.94                   & 0.93                      & 0.98                  \\
BGL              & Recall          & 0.88                   & 0.96                      & 0.98                  \\
                 & F1-Score        & 0.91                   & 0.96                      & \textbf{0.98}         \\ \midrule
                 & Precision       & 0.80                   & 0.90                      & 0.93                  \\
Thunderbird      & Recall          & 0.80                   & 0.89                      & 1.00                  \\
                 & F1-Score        & 0.80                   & 0.90                      & \textbf{0.96}         \\ \midrule
                 & Precision       & 0.94                   & 0.93                      & 0.98                  \\
Spirit           & Recall          & 0.92                   & 0.80                      & 0.96                  \\
                 & F1-Score        & 0.93                   & 0.86                      & \textbf{0.97}         \\ \bottomrule
\end{tabular}
}
\vspace{-1.5mm}
\end{table}

\subsection{RQ3: Effectiveness of \tool under different settings}
\vspace{-1mm}
\tool utilizes BERT \cite{devlin2018bert} as a pre-trained language representation to understand the semantic meaning of log messages. 
In this RQ, we would like to evaluate the performance of \tool with different pre-trained language models. We replace the BERT model in \tool with GPT2 \cite{radford2019language} and Roberta \cite{conneau2019unsupervised}, and then perform experiments to evaluate the performance of log-based anomaly detection. 
For GPT2 and Roberta encoders, we use their base model with 12 layers, 12 attention heads, and 768 hidden units. 
Table \ref{tab:RQ3_results_2} shows the results. We observe that these three pre-trained models can all understand the semantic meaning of log messages and achieve promising results. Overall, the performance of BERT is higher than that of GPT2 and Roberta. 

\begin{table}[h]
\vspace{-1.5mm}
\caption{Results of different pre-trained models}
\label{tab:RQ3_results_2}
\centering
\resizebox{0.95\columnwidth}{!}{%
\setlength{\tabcolsep}{6pt}
\renewcommand{\arraystretch}{0.9}
\begin{tabular}{@{}clccc@{}}
\toprule
\textbf{Dataset} & \textbf{Metric} & \textbf{BERT \cite{devlin2018bert}} & \textbf{GPT2 \cite{radford2019language}} & \textbf{RoBERTa \cite{conneau2019unsupervised}} \\ \midrule
                 & Precision       & 0.96          & 0.95                 & 0.85                 \\
HDFS             & Recall          & 1.00          & 1.00                 & 1.00                 \\
                 & F1-Score        & \textbf{0.98} & 0.97                 & 0.92                 \\ \midrule
                 & Precision       & 0.98          & 0.95                 & 0.95                 \\
BGL              & Recall          & 0.98          & 0.99                 & 0.90                 \\
                 & F1-Score        & \textbf{0.98} & 0.97                 & 0.93                 \\ \midrule
                 & Precision       & 0.93          & 0.85                 & 0.78                 \\
Thunderbird      & Recall          & 1.00          & 0.91                 & 1.00                 \\
                 & F1-Score        & \textbf{0.96} & 0.88                 & 0.88                 \\ \midrule
                 & Precision       & 0.98          & 0.88                 & 0.84                 \\
Spirit           & Recall          & 0.96          & 0.95                 & 0.90                 \\
                 & F1-Score        & \textbf{0.97} & 0.91                 & 0.87                 \\ \bottomrule
\end{tabular}
}
\vspace{-1mm}
\end{table}

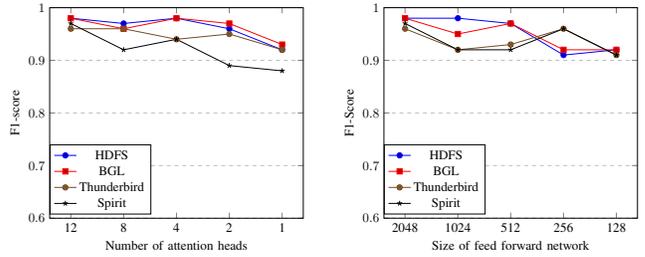
\begin{figure}[h]
    \vspace{-1mm}
    \centering
    \subfigure[Results of \tool with different number of attention heads] {
        \resizebox{0.46\columnwidth}{!}{
            \begin{tikzpicture}
            \centering
            \begin{axis}[
                ymax = 1.0,
                ymin = 0.6,
                symbolic x coords={12, 8, 4, 2, 1},
                xtick=data,
                xlabel={Number of attention heads},
                ylabel style = {
                    align=center,
                    text width=6cm
                },
                ymajorgrids=true,
                grid style=dashed,
                ylabel=F1-score,
                legend style={
                    anchor=north east,
                    at={(0.4,0.35)}
                }
            ]
            \addplot coordinates {
                (12, 0.98) (8, 0.97)  (4, 0.98)  (2, 0.96) (1, 0.92) 
            };
            \addplot coordinates {
                (12, 0.98) (8, 0.96)  (4, 0.98)  (2, 0.97) (1, 0.93) 
            };
            \addplot coordinates {
                (12, 0.96) (8, 0.96)  (4, 0.94)  (2, 0.95) (1, 0.92) 
            };
            \addplot coordinates {
                (12, 0.97) (8, 0.92)  (4, 0.94)  (2, 0.89) (1, 0.88) 
            };
            \legend{HDFS, BGL,Thunderbird, Spirit}
            \end{axis}
            \end{tikzpicture}
        }
        \label{fig:RQ4_no_heads}
    }
    \subfigure[Results of \tool with different size of feed forward network]{
        \resizebox{0.46\columnwidth}{!}{
            \begin{tikzpicture}
            \centering
            \begin{axis}[
                ymax = 1.0,
                ymin = 0.6,
                symbolic x coords={2048, 1024, 512, 256, 128},
                xtick=data,
                xlabel={Size of feed forward network},
                ylabel style = {
                    align=center,
                    text width=6cm
                },
                ymajorgrids=true,
                grid style=dashed,
                ylabel=F1-Score,
                legend style={
                    anchor=north east,
                    at={(0.4,0.35)}
                }
            ]
            \addplot coordinates {
                (2048, 0.98) (1024, 0.98)  (512, 0.97)  (256, 0.91)  (128, 0.92)
            };
            \addplot coordinates {
                (2048, 0.98) (1024, 0.95)  (512, 0.97)  (256, 0.92)  (128, 0.92)
            };
            \addplot coordinates {
                (2048, 0.96) (1024, 0.92)  (512, 0.93)  (256, 0.96)  (128, 0.91)
            };
            \addplot coordinates {
                (2048, 0.97) (1024, 0.92)  (512, 0.92)  (256, 0.96)  (128, 0.91)
            };
            \legend{HDFS, BGL,Thunderbird, Spirit}
            \end{axis}
            \end{tikzpicture}
        }
        \label{fig:RQ4_ffn_size}
    }
\caption{Results of different hyperparameter settings}
\label{fig:RQ4_results}
\vspace{-1mm}
\end{figure}

The number of attention heads and the feed-forward network size are two major hyperparameters of the Transformer model used in NeuralLog. To evaluate the impact of these parameters on detection accuracy, we vary their values and perform experiments on the four datasets. The resulting F1-scores are shown in Figure \ref{fig:RQ4_results}.
We observe that reducing the number of attention heads and feed-forward network size can slightly hurt the performance of NeuralLog. For example, \tool achieves F1-scores ranging from 0.96 to 0.98 when using 12 attention heads. These results are higher than those obtained by using only one attention head (0.88 - 0.93). Similarly, F1-scores achieved by a larger feed-forward network are usually better.
Overall, the Transformer model achieves promising results with different hyperparameter values (most F1-scores are above 0.90). We observe that the performance is best when the number of attention heads is between 4 and 12, and the feed-forward network size is from 512 to 2048.


\vspace{-1mm}
\section{Discussion}
\label{sec:discussion}
\subsection{Why does \tool Work?}
There are three main reasons that make \tool perform better than the related approaches. First, \tool directly uses raw log messages instead of using a log parser in preprocessing. Since there is no loss of information from log messages, \tool can precisely learn the semantic representation of log messages, compared to other approaches that depend on log parsing. Second, \tool leverages BERT \cite{devlin2018bert} and WordPiece \cite{schuster2012japanese, wu2016google} to capture the meaning of OOV words at the subword level. Moreover, the transformer-based classification model can also improve the performance of anomaly detection. The transformer utilized by \tool can learn different sequence patterns in log messages and determine which patterns are more relevant to anomalies.


Our study has demonstrated the effectiveness of \tool for anomaly detection. However, \tool still has limitations. 
Our approach is based on the learning of the semantic meanings of log messages. Given a log message, we first remove those words that contain numbers and special characters. However, in some cases, the removed words may carry important information, such as node ID, task ID, IP address, or exit code. These information could be useful for anomaly detection in certain scenarios. In our future work, we will encode more log-related information and investigate their impact on log-based anomaly detection.

\subsection{Threats to Validity}

We have identified the following major threats to validity.

\textbf{Subject datasets}. In this work, we use datasets collected from the distributed system (i.e., HDFS) and supercomputer (including BGL, Thunderbird, and Spirit). Although these datasets all come from real-world systems and contain millions of logs, the number of subject systems is still limited and do not cover all the domains. 
In the future, we will evaluate the proposed approach on more datasets collected from a wide variety of systems.


\textbf{Tool comparison}. In our evaluation, we compared our results with those of related approaches (i.e., SVM, LR, IM, LogRobust, and Log2Vec). 
We adopt the implementation of SVM, LR, and IM-based methods provided by Loglizer \cite{loglizer}. We adopt the implementation of LogRobust and Log2Vec provided by their authors. 
We apply the default parameters and settings (e.g., sliding window size, step size, etc.) used in the previous work \cite{he2016experience, meng2020semantic, zhang2019robust}. Still, the correctness of these implementations could be a threat.
To reduce this threat, we make sure that the implementation of related work can produce similar results as those reported in the original papers.

\textbf{Noises in labeling}. Our experiments are based on four public datasets that are widely used by related work \cite{li2020_swisslog, zhang2020anomaly, meng2019loganomaly, nedelkoski2020self, he2016experience}. These datasets are manually inspected and labeled by engineers. Therefore, data noise (false positive/negatives) may be introduced during the manual labeling process. Although we believe the amount of noise is small (if it exists), we will investigate the data quality issue in our future work. 

\vspace{-1mm}
\section{Related Work}
\label{sec:related}

\subsection{Log Parsing Errors}

The log parsing accuracy highly influences the performance of log mining \cite{he2016evaluation}. Log parsers could produce inconsistent results depend on the preprocessing step and the set of parameters \cite{he2016evaluation, zhu2019tools}. The preprocessing step can further improve log parsing accuracy \cite{he2016evaluation} and despite the simplicity, it still requires some additional manual work \cite{zhu2019tools}.
Zhu et al. \cite{zhu2019tools} benchmarked 13 automated log parsers on a total of 16 datasets. They found that Drain \cite{he2017drain} is the most accurate log parser, which attains high accuracy on 9 out of 16 datasets. The other top-ranked log parsers include IPLoM \cite{makanju2009clustering}, AEL \cite{jiang2008abstracting} and Spell \cite{du2016spell}. They also found that some model parameters need to be tuned manually, and some models did not scale well with the volume of logs. He et al \cite{he2016evaluation} evaluated four widely used log parsers, including SLCT \cite{vaarandi2003data}, IPLoM \cite{makanju2009clustering}, LKE \cite{fu2009execution} and LogSig \cite{tang2011logsig}.

In practice, new types of logs always appear \cite{meng2019loganomaly}, as OOV words can be added to log templates and lead to many extra log events, which will confuse the downstream tasks.
Zhang et al. \cite{zhang2019robust} indicated that log data is unstable, meaning that new log events often appear due to software evolution at its lifetime. Their empirical study on a Microsoft online service system shows that up to 30.3\% logs are changed in the latest version. 
In our work, we perform an empirical study of the log parsing errors caused by the OOV problem and semantic misunderstanding, and investigate their impact on the performance of anomaly detection.


\subsection{Log Representations}
As described in Section \ref{sec:motivation}, most of the existing log-based anomaly detection approaches use log parsers to obtain log events and represent log messages as log events.
Therefore, the existing approaches suffer from the OOV problem and the inaccurate log parsing.
Recently, deep learning-based models have been adopted into log-based anomaly detection. DeepLog \cite{du2017deeplog} applies Spell \cite{du2016spell} to extract log events, then each log event is assigned with an index. Since DeepLog represents log messages as the indexes of log templates, it cannot prevent semantic information loss and could produce many wrong detection results \cite{li2020_swisslog, meng2020semantic}.
LogRobust \cite{zhang2019robust} leverages Drain \cite{he2017drain} to obtain log templates, then encodes these templates using the FastText \cite{joulin2016fasttext} framework combined with TF-IDF \cite{salton1988term} weight. 
LogAnomaly \cite{meng2019loganomaly} applies FT-Tree \cite{zhang2017syslog} to parse log messages to templates, then proposes template2Vec to encode these templates based on Word2Vec \cite{le2014distributed}.
SwissLog \cite{li2020_swisslog} 
obtains the semantic information of log messages after parsing log messages using a dictionary-based approach. Due to imperfect log parsing, these methods could fail to capture the semantic meaning of log messages and produce incorrect results.
Log2Vec \cite{meng2020semantic} transforms raw log messages into semantic vectors. 
As it utilizes character-level features, it could not effectively handle some domain-specific words \cite{sasaki2019subword, zhao2018generalizing, hu2019few}. 
Besides, Log2Vec adopts word2vec-based model that ignores the contextual information in sentences \cite{li2020_swisslog}, thus it cannot fully understand the semantic meaning of log messages.
Nedelkoski et al. \cite{nedelkoski2020self} proposed Logsy, which is a classification-based method to learn log representations in a way to distinguish between
normal data from the target system and anomaly samples from auxiliary log datasets. It does not provide mechanism for handling OOV words in log messages either.

To overcome the limitations of existing methods, we propose NeuralLog, a deep learning-based anomaly detection approach using raw log data. 
NeuralLog utilizes WordPiece tokenization to effectively handle OOV words that constantly appear in log messages. It also leverages BERT, a widely used pre-trained language representation, to understand the semantic meaning and capture the contextual information of raw log messages. 
Combined with a Transformer-based classification model, NeuralLog achieves high accuracy on anomaly detection. 
Furthermore, we only use log data from the target systems and do not require any auxiliary data.

\vspace{-1mm}
\section{Conclusion}
\label{sec:conclusion}
Log-based anomaly detection is important for improving the availability and reliability of large-scale software systems. Our empirical study shows that existing approaches suffer from inaccurate log parsing and cannot handle OOV words well. 
To overcome the limitations introduced by log parsing, in this paper, we propose NeuralLog, 
a log-based anomaly detection approach that does not require log parsing. Our approach employs BERT encoder to capture the semantic meaning of raw log messages. To better capture contextual information from log sequences, we construct a Transformer-based classification model. 
We have evaluated the proposed approach using four public datasets. The experimental results show that \tool is effective and efficient for log-based anomaly detection.

Our source code and experimental data are publicly available at \url{\repourl}.



\section*{Acknowledgment}

This research was supported by the Australian Government through the Australian Research Council's Discovery Projects funding scheme (project DP200102940).

\vspace{-1mm}

\bibliographystyle{IEEEtran}
\bibliography{sample}
\balance

\end{document}